\documentclass[superscriptaddress,aps,prl,twocolumn]{revtex4}

\usepackage{graphicx}
\usepackage{amsmath}
\usepackage{amsfonts}
\usepackage{amssymb}
\usepackage{epsf}
\usepackage{hyperref}

\begin{document}
\title{From multimode to monomode guided atom lasers: an entropic analysis}
\author{G. L. Gattobigio}
\affiliation{Laboratoire Kastler Brossel, Ecole Normale Sup\'erieure,
24 rue Lhomond, 75005 Paris, France}
\affiliation{Laboratoire de Collisions Agr\'egats R\'eactivit\'e,
CNRS UMR 5589, IRSAMC, Universit\'e Paul Sabatier, 118 Route de
Narbonne, 31062 Toulouse CEDEX 4, France}
\author{A. Couvert}
\affiliation{Laboratoire Kastler Brossel, Ecole Normale
Sup\'erieure, 24 rue Lhomond, 75005 Paris, France}
\author{M. Jeppesen}
\affiliation{Laboratoire Kastler Brossel, Ecole Normale
Sup\'erieure, 24 rue Lhomond, 75005 Paris, France}
\affiliation{Australian Centre for Quantum Atom Optics, Physics
Department, The Australian National University, Canberra, 0200,
Australia}
\author{R. Mathevet}
\affiliation{Laboratoire de Collisions Agr\'egats R\'eactivit\'e,
CNRS UMR 5589, IRSAMC, Universit\'e Paul Sabatier, 118 Route de
Narbonne, 31062 Toulouse CEDEX 4, France}
\author{ D. Gu\'ery-Odelin}
\affiliation{Laboratoire Kastler Brossel, Ecole Normale
Sup\'erieure, 24 rue Lhomond, 75005 Paris, France}
\affiliation{Laboratoire de Collisions Agr\'egats R\'eactivit\'e,
CNRS UMR 5589, IRSAMC, Universit\'e Paul Sabatier, 118 Route de
Narbonne, 31062 Toulouse CEDEX 4, France}
 \date{\today}
\begin{abstract}
We have experimentally demonstrated  a high level of control of the
mode populations of guided atom lasers (GALs) by showing that the entropy per particle
of an optically GAL, and the one of the trapped Bose
Einstein condensate (BEC) from which it has been produced are the
same. The BEC is prepared in a crossed beam optical dipole trap. We have achieved isentropic outcoupling for both magnetic and optical schemes. We can prepare GAL
in a nearly pure monomode regime (85 \% in the ground state). Furthermore, optical outcoupling
enables the production of spinor guided atom lasers and opens the possibility to tailor their polarization.
\end{abstract}
\pacs{???}

\maketitle

Isentropic transformations have been used extensively to
manipulate classical and degenerate quantum gases. For instance, the
reversible formation of a molecular BEC from ultra-cold fermionic
atoms having two spin components was performed by adiabatically tuning the
inter-species scattering length from positive to negative values
\cite{CSC04,WNC04,BKC04,PSK05,ZSS05}. Another illustration is the
adiabatic change of the shape of the confining trap of cold atoms giving the possibility
to change the phase space density in a
controlled manner \cite{PMW97, SMC98}. A spectacular demonstration of this idea has
been the multiple reversible formation of BEC by adiabatically superimposing
an optical dimple trap on a magnetically trapped and
pre-cooled sample of atoms~\cite{SMC98}. For an ideal transformation,
the entropy of the initial cloud should remain constant. However, the limitations of the
experimental setup always introduce an extra source of entropy.
The challenge for the experimentalists is to minimize this latter contribution.

In this experiment, we demonstrate the
control and the characterization of a guided atom laser (GAL)
\cite{GRG06,CJK08}. The experiments performed to date on the beam
quality were addressing the spatial mode of free-falling atom
lasers. The importance of the outcoupling scheme or the role
played by atom-atom interactions has been extensively studied
\cite{BHE99,RGC06, JDD08}. GALs are characterized by
the population of the transverse modes of the guide. We have extended the use of
isentropic analysis to propagating matter waves in order to relate
quantitatively these populations to the characteristics of the BEC
from which the GAL originates. This approach turns out to be
possible because of both the validity of the local thermal equilibrium
and the sufficiently large reduction of the extra entropy production
generated by the experimental manipulation. Improvements to the
production and characterization of the GAL are crucial for
fundamental studies such as quantum transport \cite{transport},
and applications in metrology \cite{metrology}, among others.

\begin{figure}[t!]
    \begin{center}
        \includegraphics[width=9cm]{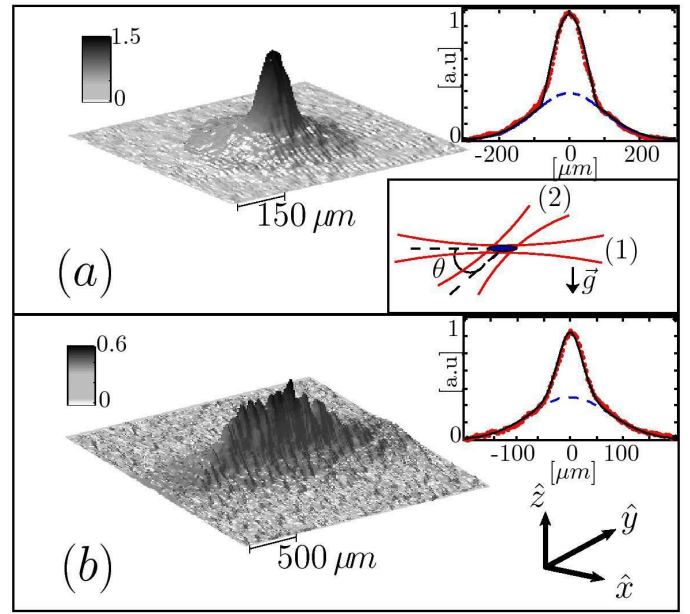}
    \end{center}
    \caption{(Color online). $(a)$ Absorption image of a BEC of $^{87}$Rb atoms released from a
crossed dipole trap, after 20 ms time-of-flight. Inset top right:
Atom density across a line through the cloud center. Inset bottom
right: Crossed dipole trap laser geometry. $(b)$ Absorption image
of a guided atom laser, outcoupled from the same BEC and imaged in
the same conditions. The atom laser has mean excitation number in
the transverse modes of $\langle n \rangle = 4$. Inset: atom
density integrated along the guide axis (1). A bimodal density
structure is evident, as in $(a)$ above.}
    \label{fig1}
\end{figure}

The experiment starts by loading $3\times 10^7$ $^{87}$Rb-atoms in
a crossed beam optical dipole trap at a wavelength of 1070 nm from an elongated magneto-optical trap
(MOT). To transfer the atoms in the lower hyperfine level $F=1$,
we align the horizontal arm (1) of the optical trap with the long
$\hat{y}$-axis of the MOT (see inset of Fig.~\ref{fig1}.$a$) and
we mask the repump light in the overlapping region between the two
traps. The other arm (2) of the crossed dipole trap makes an angle
of $\theta = 45^{\circ}$, in the $\hat{y}\!-\!\hat{z}$ plane where the
$\hat{z}$ axis is the vertical axis \cite{CJK08}. In practice, the beam (1) (resp. (2)) has
a waist of $w_{\rm 1}\,=\, 40\,\mu\rm{m}$ (resp. $w_{\rm
2}\,=\,130\,\mu\rm{m}$) and an initial power of $P_{\rm 1}\,\simeq
\,20\,W$ (resp. $P_{\rm 2}\,\simeq\,100\,W$).

\begin{figure}[t!]
    \begin{center}
        \includegraphics[width=9cm]{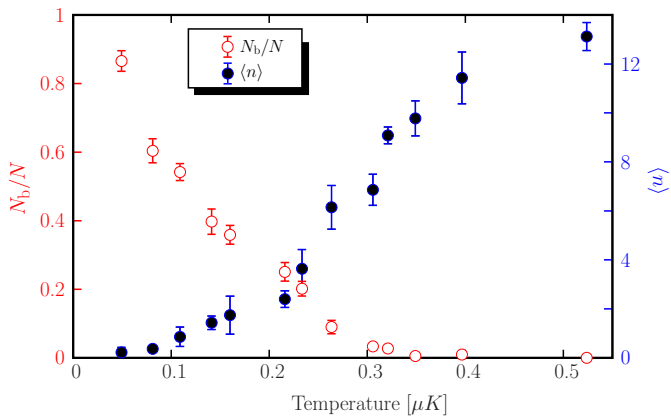}
    \end{center}
    \caption{(Color online) : Measured condensed fraction $N_{\rm b}/N$ (open circles) of the source BEC and mean excitation number,  $\langle n \rangle$,
    (solid circles) of the guided
atom laser produced from the source BEC by magnetic outcoupling.
Both are shown as a function of the temperature of the trapped
source BEC immediately prior to the outcoupling.}
    \label{fig2}
\end{figure}

The BEC transition temperature is on the
order of 300 nK. Typically, a BEC of
$2\times 10^{5}$ atoms is formed after a $4\,{\rm s}$ evaporation ramp carried out by lowering the power of
both beams (see Fig.~\ref{fig2}). Using the spin-distillation technique during the evaporation process \cite{CJK08},
we prepare the BEC in a purely $F=1,m_{\rm F}=0$ state.
The density profile distribution of the BEC is obtained from absorption images after a 20 ms
time-of-flight (TOF). Fig.~\ref{fig1}.$a$ shows a typical BEC profile with a significant
thermal fraction.

We first describe experiments performed using magnetic
outcoupling. At the desired temperature, we prepare the cloud by increasing the power of the two beams in
$200\,{\rm ms}$ to limit residual evaporation. Beam (1) is used as
an horizontal guide whose frequency
$\displaystyle{\omega_{\perp}/2\,\pi = 289 \,{\rm Hz}}$ is
measured by exciting the transverse sloshing mode. The power of
the beam (2) is then reduced to facilitate the outcoupling, while
the power of the beam (1) is kept constant (inset of
Fig.~\ref{fig1}.$a$). The outcoupling is performed by applying a
magnetic gradient with increasing strength as a function of time.
During this $100\,{\rm ms}$ outcoupling phase, outcoupled atoms
propagate in the optical guide over 1 mm before being
imaged. Fig.~\ref{fig1}.$b$ shows a density profile of the 
GAL generated from the BEC of Fig.~\ref{fig1}.$a$. 
Such pairs of images have been acquired for different
temperatures of the atomic cloud.

To relate quantitatively the characteristics of the GAL to those
of the cloud it originated from, we model the GAL using a
thermodynamic approach within the grand-canonical ensemble
\cite{CNY96,Yvan,CJK08}. We consider an ideal Bose
gas enclosed in a fictitious box along the longitudinal axis, and
confined transversally by a harmonic potential. For a given linear
density $\rho$, a transverse condensation occurs below a critical
temperature $T^{c}_a$, i.e. a macroscopic fraction of the atoms
occupy the transverse ground state independently of their
longitudinal state \cite{tca}. Indeed, there is no Bose-Einstein
condensation in the true ground state (transverse and
longitudinal) of this quasi one-dimensional system in the
thermodynamic limit \cite{Yvan}.

The transverse profile (integrated along the $\hat y$ axis) is
readily inferred from the thermodynamic model:
\begin{equation}
\lambda\rho(x) = \frac{g_{1/2}(z_a)}{\sqrt{\pi \sigma^2}}
\text{e}^{-x^2/\sigma^2} + \frac{1}{a_0\sqrt{2\pi\xi_a^3}}
g_2\left(z_a\text{e}^{-\xi_a x^2/2a_0^2}\right).
\end{equation}
where $a_0=(\hbar/m\omega_\perp)^{1/2}$ is the transverse harmonic
oscillator length, $\sigma$
 is the size of the ground state \cite{inter}, $z_a=\exp(\beta\mu)$ the fugacity, $\xi_a=\beta
\hbar\omega_\perp$ with $\beta=1/k_BT$  \cite{index},
$\lambda=h/(2\pi m k_BT)^{1/2}$ and $g_{p}(z)=\sum_{n=1}^\infty
z^n/n^{p}$ the Bose function of index $p$.

To extract quantitative data from our profiles, we have
approximated the $g_2$ function by a Gaussian and fit our data
with the sum of two Gaussians. We have checked numerically that
this method does not introduce significant errors with respect to
experimental uncertainties. In this way, we can calculate the
fraction of condensed atoms, the temperature, and we can thus
obtain the mean excitation number through the formula
\cite{expl}:
\begin{equation}
\langle n \rangle= \left(1-\frac{\rho_0}{\rho}\right)\langle n
\rangle_{\rm th},
    \label{equation}
\end{equation}
where $\rho_0=g_{1/2}(z_a)/\lambda$ is the mean linear density in
the transverse ground state and $\langle n \rangle_{\rm
th}$ is the average excitation number of
the excited atoms which is a function of $\xi_a$. The first order
expansion $\langle n \rangle_{\rm
th}\simeq \xi_a^{-1}-1/2$ gives a satisfactory estimate over the whole range of
parameters that we have used.

In practice, the two measurements of the linear density and of
$\langle n \rangle$ give access to the two parameters $\xi_a$ and
$z_a$, and therefore characterize completely the thermodynamic
equilibrium quantities. We have plotted in Fig.~\ref{fig2} the
measured value for $\langle n \rangle$ along with the condensed
fraction of the source cloud of atoms from which the GAL has been
produced, as a function of the temperature of the cloud. There is
a clear correlation between the two quantities: the colder the
sample, the more monomode the guided atom laser. The mean number
of populated energy levels, taking into account their degeneracy,
varies from more than $\sim$ 100 levels for the atom laser
extracted from a cloud at 500 nK, to just one level populated (for
the coldest point at 50 nK, 85\% of atoms are in the transverse
ground state of the guide).

In the following, we compare quantitatively the entropy per atom
in both systems. This comparison is carried out in a range of
temperature for the source cloud such that $\hbar\omega \leq gn_b
\leq k_BT < 0.9 k_BT^c_b$, where $T^c_b$ is the BEC transition
temperature, $n_b$ is the atomic density of the condensate and
$g=4\pi\hbar^2a/m$ is the strength of the interaction.
In this temperature regime, Hartree-Fock and Popov theories yield
very similar predictions for thermodynamic quantities which are well reproduced by the ideal gas formula
\cite{GPS97,CSC04}. For our GAL, the interactions are such that
$\rho_0 a < 0.5$, which validates the use of a non interacting
model described above.

We have therefore computed independently the entropy per particle $S/k_BN$ using
the non interacting formula for both systems. This quantity is
intensive and, in both cases, depends only on the two
dimensionless and intensive parameters: the fugacity $z$ and
$\xi=\beta \hbar \omega$. Within our thermodynamic model, the entropy per particle of the GAL reads
\begin{eqnarray}
& & \frac{S_{a}(z_a,\xi_a)}{k_BN_a}   =   -\log z_a+
\rho\lambda^{-1}\sum_{p=1}^\infty \bigg[ \frac{z_a^p}{p^{3/2}}
\frac{1}{(1-e^{-p_a\xi_a})^{2}}  \nonumber \\
& &
+ \frac{p+1}{2}g_{3/2}(z_ae^{-p\xi_a}) + \frac{2\xi_a
z_a^p  e^{-p\xi_a}}{p^{1/2}} \frac{1}{(1-e^{-p\xi_a})^{3}} \bigg].
\label{eq3}
\end{eqnarray}

\begin{figure}[t!]
    \begin{center}
        \includegraphics[width=8.5cm]{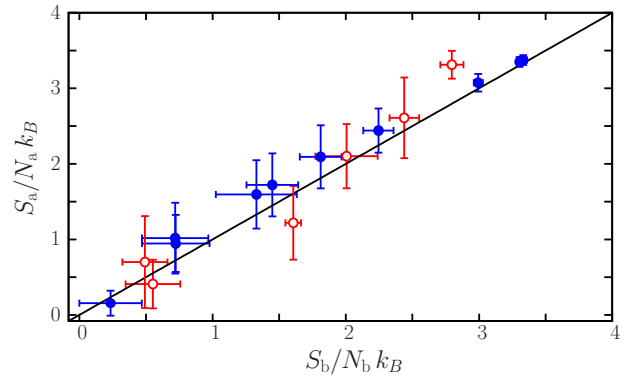}
    \end{center}
    \caption{(Color online). Entropy per atom of the atom laser
$(S_{\rm a}/N_{\rm a} k_{\mathrm{B}})$ against the entropy per
atom of the source BEC  $(S_{\rm b}/N_{\rm b}
    k_{\mathrm{B}})$. Each point is the average
of ten measurements. The different points correspond to different
initial temperatures of the BEC. Solid circles are for magnetic
outcoupling, open circles for optical outcoupling. The solid line
is the expected result for equal entropy per particle of both
systems.}
    \label{fig3}
\end{figure}

Figure \ref{fig3} shows experimental measurements of the entropy per particle of the source BEC compared to the corresponding GAL. The solid line represents perfect isentropic outcoupling (that is, $S_{\rm a}/N_{\rm a}k_B=S_{\rm b}/N_{\rm b}k_B$). Our
data (solid circles) are therefore in good agreement with the
isentropic assumption used to model the outcoupling.

The robustness of the entropic analysis has been confirmed by
studying the GAL generated by optical outcoupling (i.e. outcoupling by reducing the power in laser beam (2), with no magnetic field used). After the
evaporation ramp and subsequent recompression of the crossed
dipole trap (as with magnetic outcoupling), the outcoupling is performed by decreasing  the intensity of beam (2) progressively until atoms spill into the
optical guide formed by beam (1) (see Fig.~\ref{fig1}). The outcoupled atoms
experience a force due both to
gravity (since beam (1) is not perfectly horizontal) and the local dipole force (since beam (2) does not
necessarily cross beam (1) at its waist position). The comparison
between the entropies of the atom laser and the BEC from which
it has been generated are plotted in Fig.~\ref{fig3} (open circles) which   confirms that this optical scheme is also nearly isentropic.
Conversely, one can predict the
characteristics of the guided atom laser by knowing those of the
condensate from which it has been outcoupled  (assuming an
isentropic transformation).

The control of the mean number of excitations $\langle n \rangle$
could be used to explore the transition between the classical
regime, in which a large number of energy levels are populated, to
a pure quantum regime where nearly all atoms are in the transverse
ground state of the confinement. For instance, this system could
be well suited to investigate classical versus quantum chaos where
sigificant differences are expected between the two regimes
\cite{chaos}.

The model used to deduce the entropy assumes that the GAL is at
local thermodynamic equilibrium. The collision rate evaluated in
the moving frame and using the formula of the classical limit is
on the order of 25 s$^{-1}$ for typical experimental parameters.
It is strongly enhanced because of bosonic amplification
\cite{JGZ97}. During the 100 ms of the outcoupling process, the
local thermal equilibrium assumption is thus valid. We have also
performed outcoupling over shorter times (50 ms) and obtained
similar agreement for the entropy per particle.

As the GAL propagates, its density is diluted due to velocity
dispersion and inhomogeneous forces on the atoms. The propagation
is therefore accompanied by a modification of the transverse mode
population, as soon as the collision rate is sufficient to
validate the local equilibrium assumption. This is a significant
difference with light propagation in fibers. The expected
modification for a 1 mm propagation length is compatible with our
error bars when we compare the results of long outcoupling time
with respect to short outcoupling time. After a sufficiently long
propagation, the mode population is expected to freeze out since
the atomic density decreases and the collision rate becomes too
small for restoring the thermal equilibrium. In this limit, the
thermodynamic model no longer applies.

\begin{figure}[t!]
    \begin{center}
        \includegraphics[width=8.5cm]{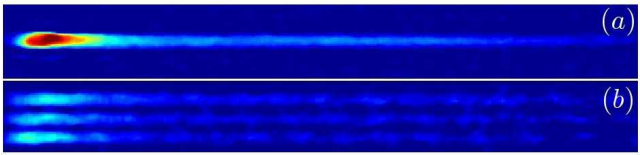}
    \end{center}
    \caption{(Color online). (a)
Spinor guided atom laser obtained by optically outcoupling atoms
from a spinor condensate. (b) The three components $m_F=-1,0,+1$
are separated by applying a Stern and Gerlach field during the
time-of-flight.
    }
    \label{fig4}
\end{figure}

The optical outcoupling process affects atoms of all Zeeman states equally. Also, the optical waveguide confines all Zeeman states equally, compared to a magnetic guides in which only certain Zeeman states may propagate. It is therefore possible to produce atom lasers that contain mixtures of atoms in different spin states, or even linear superpositions of spin states.  Starting with a spinor
BEC (see \cite{spinor}  and references therein) in a incoherent
mixture of $m_F=\pm 1,0$, we have been able to generate a spinor
guided atom laser as shown in Fig.~\ref{fig4} where the three
components are separated using a Stern and Gerlach field during
the TOF.

As a preliminary study, we have produced a BEC in superposition of states of the $F=1$ and
$F=2$ hyperfine levels, by combining the spin distillation technique \cite{CJK08}, and microwave pulses
with well-defined polarization \cite{MHJ98}. The internal degrees of freedom of the atoms are the
analogue of polarization for light. The dimension of the
corresponding subspace is much larger for atoms than for light.
An appealing perspective, which requires a well-controlled magnetic environment,
 lies in the full control of the polarization of the guided atom laser over its propagation.

In this article, we have demonstrated an unprecedented control
over the transverse degrees of freedom of a guided atom laser. 
Two different outcoupling schemes are analyzed by a quantitative model based on an entropy analysis. In addition, optical
outcoupling offers the possibility to control the internal degrees
of freedom. A related prospect is the generation of a GAL with a
very narrow longitudinal velocity dispersion. This is in principle
achievable by exploiting the tunnel effect for the outcoupling
\cite{DMR06}.

Another promising perspective of the techniques developed  in this
article is the production of a continuous guided atom laser. The
periodic replenishing of Bose Einstein condensates held in an
optical trap has already  been demonstrated in Ref.~\cite{CSL02}.
Using the spin distillation technique \cite{CJK08}, this reservoir
of condensates could be produced in a $m_F=0$ state. Applying a
gradient of magnetic field along one arm of the dipole trap, atoms
could be continuously extracted in it by applying a
radio-frequency to transfer atoms into a magnetically sensitive
state $m_F=\pm 1$.

\begin{acknowledgments}
We acknowledge fruitful discussions with T. Lahaye, Y. Castin and J. Dalibard, and financial support from the D\'el\'egation
G\'en\'erale pour l'Armement, the Institut Francilien de Recherche
sur les Atomes Froids (IFRAF) and the Plan-Pluri Formation (PPF).
\end{acknowledgments}


\begin{thebibliography}{99}


\bibitem{CSC04} L. D. Carr, G. V. Shlyapnikov, and Y. Castin, Phys. Rev. Lett. \textbf{92}, 150404 (2004).
\bibitem{WNC04} J. E. Williams, N. Nygaard, and C.W. Clark, New J. Phys. \textbf{6}, 123 (2004).
\bibitem{BKC04} T. Bourdel,  \emph{et al.} Phys. Rev. Lett. \textbf{93} 050401 (2004).

\bibitem{PSK05} G. B. Partridge, \emph{et al.} Phys. Rev. Lett. \textbf{95} 020404 (2005).
\bibitem{ZSS05} M. W. Zwierlein,  \emph{et al.}, Phys. Rev. Lett. \textbf{94} 180401 (2005).

\bibitem{PMW97} P. W. H. Pinkse, \emph{et al.} Phys. Rev. Lett. \textbf{78}, 990 (1997).
\bibitem{SMC98} D. M. Stamper-Kurn,  \emph{et al.} Phys. Rev. Lett. \textbf{81}, 2194 (1998).

\bibitem{GRG06} W. Guerin, \emph{et al.} Phys. Rev. Lett. \textbf{97}, 200402 (2006).
\bibitem{CJK08} A. Couvert, \emph{et al.}, EPL \textbf{83}, 50001 (2008).

\bibitem{BHE99} I. Bloch, T. W. H\"ansch, and T. Esslinger, Phys. Rev. Lett. \textbf{82}, 3008 (1999).
\bibitem{RGC06} J.-F. Riou,  \emph{et al.} Phys. Rev. Lett. \textbf{96}, 070404 (2006).
\bibitem{JDD08} M. Jeppesen, \emph{et al.} Phys. Rev. A \textbf{77}, 063618 (2008).

\bibitem{transport}
P. Leboeuf and N. Pavloff, Phys. Rev. A \textbf{64}, 033602 (2001);
I. Carusotto and G.~C. La~Rocca, Phys. Rev. Lett.  \textbf{84}, 399 (2000);
T. Paul, K. Richter and P. Schlagheck, Phys. Rev. Lett. \textbf{94}, 020404 (2005).
\bibitem{metrology}  \emph{Proceedings of the 6th Symposium on Frequency
  Standards and Metrology} (Ed. P. Gill, World Scientific, 2002).

\bibitem{CNY96} T. T. Chou, C. N. Yang, and L. H. Yu, Phys. Rev. A \textbf{53}, 4257 (1996).
\bibitem{Yvan} E. Mandonnet, \emph{et al.}, Eur. Phys. J. D \textbf{10}, 9 (2000); Y. Castin,  \emph{et al.} J. Mod. Opt. \textbf{47} 2671 (2000).

\bibitem{tca} $k_BT^c_a = \hbar \omega_\perp ( \rho \lambda_c/\zeta(5/2))^{1/2}$,
with $\zeta(5/2)\simeq 1.34$ and $\lambda_c=h/(2\pi m
k_BT^{c}_a)^{1/2}$. This semiclassical expression is valid for our parameters, since, for the lowest
measured temperature, we found $k_B T \simeq 4 \hbar\omega_\perp$.

\bibitem{inter} A two dimensional Gaussian ansatz for
the transverse degrees of freedom inserts in the Gross-Pitaevskii
energy functional yields the size of
the ground state for our model $\sigma=a_0(1+2\rho_0a)^{1/4}$, where $a$ is the
scattering length and $\rho_0$ the linear density of the ground state.

\bibitem{index} We use the index $_b$ for the BEC cloud and $_a$
for the guided atom laser.

\bibitem{expl} $\langle
n\rangle=\sum_{\mathbf{n}}\left(n_{x}+n_{y}\right)\pi_{\mathbf{n}}/2$
where $\pi_{\mathbf{n}}$ is the population of the state with
quantum numbers $(n_{x},n_{y})$. Therefore $\langle
n\rangle=\left(1-N_{0}/N\right)\sum_{\mathbf{n}\neq(0,0)}\left(n_{x}+n_{y}\right)\pi^{\rm
th}_{\mathbf{n}}/2=\left(1-N_{0}/N\right)\langle n \rangle_{\rm
th}$ with $\pi^{\rm th}_{\mathbf{n}}=\pi_{\mathbf{n}}N_0/(N-N_0)$
and $\pi^{\rm th}_{0}=0$.

\bibitem{GPS97} S. Giorgini, L. P. Pitaevskii, and S. Stringari, Phys. Rev. Lett. \textbf{78} 3987 (1997).



\bibitem{chaos} \emph{Quantum versus chaos: questions emerging from mesoscopic cosmos} by Katsuhiro Nakamura, (Kluwer Academic Publishers, New York, 2002).

\bibitem{JGZ97} D. Jaksch, C. W. Gardiner, and P. Zoller, Phys. Rev. A \textbf{56}, 575 (1997).

\bibitem{spinor} D.M. Stamper-Kurn and W. Ketterle, ``Coherent Atomic Matter Waves'', Les Houches
Summer School Session LXXII in 1999, edited by R. Kaiser, C.
Westbrook, and F. David (Springer, NewYork, 2001).

\bibitem{MHJ98}  
M. R. Matthews, \emph{et al.} Phys. Rev. Lett. \textbf{81}, 243 (1998).

\bibitem{DMR06} F. Delgado, J. G. Muga, and A. Ruschhaupt, Phys. Rev. A \textbf{74}, 063618 (2006).
\bibitem{CSL02} A. P. Chikkatur {\it et al.}, Science \textbf{296}, 2193 (2002).


\end{thebibliography}
\end{document}